# Effect of Feedback on Oscillation Characteristics of a Device with Virtual Cathode

A. E. Hramov

**Abstract**—The effect of external feedback on the oscillation characteristics of a device with the virtual cathode (VC) is investigated by the method of numerical modeling. A strong influence of the feedback phase delay on the frequency and power level of signals generated in the system is demonstrated.

Millimeter-wave devices operating on the basis of the virtual cathode (VC) with electron beams attract increasing interest as promising microwave-power sources in power engineering, communications, and radiolocation [1, 2]. In this connection, the necessity arises of controlling both the amplitude and frequency characteristics of the radiation generated in such systems. In the future, it may lead to the possibility of creating super-powerful systems comprising several vircators that operate in parallel [3–5].

Recently, various VC devices with different types of feedback have been proposed (see, for example, [6, 7]) to control the characteristics of the generated radiation in the vircator system. Among these devices, the virtod [6], or an oscillator using external feedback with overcritical current, is of special interest. The feedback in the virtod is implemented by withdrawing approximately 15% of the oscillation power from the region of the VC formation to the region, in which the relativistic electron beam is accelerated along the external phase-delay line.

In this paper, the effect of the delay on the virtod oscillation characteristics is investigated using numerical simulation based on solution to the Maxwell–Vlasov self-consistent system of equations [8, 9].

The system under consideration is a cylindrical waveguide section. A large-current annular single-velocity electron beam with the relativistic factor $\gamma_0 = 2.7$ is injected into the system. A sufficiently strong axial magnetic field is applied along the system axis, so that the beam electrons are completely guided by this field. The characteristic feature of the system is the external feedback caused by kinematic modulation of the electrons entering the drift space by the electromagnetic signal generated in the interaction space (in the VC region) that influences the beam with the phase delay $\beta$.

We investigated the system characteristics for a fixed value of the dimensionless current $\alpha = 2$, where $\alpha$ is the ratio of the beam current to the maximum vacuum current. Note that the given value of the current corresponds to the maximal efficiency of the system without the feedback (see, for example, [10, 11]).

Figure 1 shows the dependence of the electron interaction power in the system under study on the feedback-circuit phase delay $\beta$ and the curve obtained in [6] that represents the results of the experiment on location. We see a sufficiently good qualitative agreement between the results of simulation and physical experiment. The horizontal line corresponds to the power level in the system without the feedback. For the optimal $\beta$, the efficiency of power exchange in the system increases by a factor of 1.5 or more.

Variation of the phase delay $\beta$ when the current is fixed yields also a readjustment of the oscillation frequency within sufficiently broad limits (Fig. 2). This can be illustrated using the simplest phenomenological model. In [12], the phase synchronization of the microwave VC oscillator by the external signal is described by a model based on the Van der Pol equation. In our case, in order to describe the stabilization of the oscillator by its intrinsic signal that influences upon the system after passing through the phase-delay circuit, we can represent this model in the form

$$\frac{d^2E}{dt^2} - 2\gamma\left(1 - \frac{E_0^2}{E_{NL}^2}\right)\frac{dE}{dt} + \omega_0^2 E = \frac{2\omega_0^2}{Q} E(\omega_0 t - \beta). \quad (1)$$

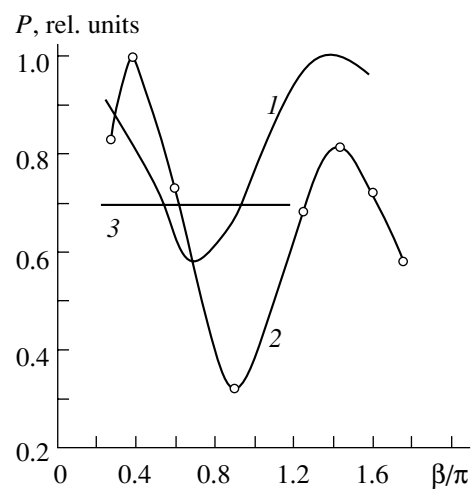

**Fig. 1.** Dependence of the oscillation power $P$ on the feedback phase: (*1*) calculations; (*2*) experiment on location (the data borrowed from [6]); and (*3*) the power level of the free VC oscillations.

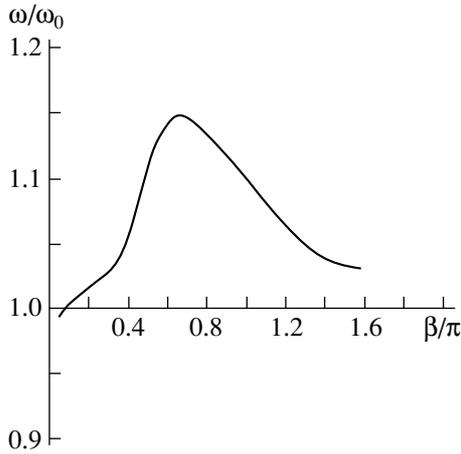

**Fig. 2.** Dependence of the oscillation frequency on the feedback-circuit phase delay.

Here, $E$ is the field in the system; $E_0$ is the field amplitude; $E_{NL}$ is the value of the nonlinear upper boundary of the field amplitude; $\gamma$ is the oscillation increment; and $\omega_0$ and $Q$ are the cold resonance frequency and $Q$-factor of the resonance system, respectively.

In order to solve (1) in terms of the amplitude and phase, we assume that $E$ has the form $E = E_0(t)\cos(\omega_0 t - \phi(t))$ ($E_0$ and $\phi$ are slowly varying amplitudes). Applying the Van der Pol averaging method, one can show that, in the first approximation (i.e., under the assumption that $E_0(t) = $ const), the dynamics of the phase $\phi(t)$ is described by the equation

$$\frac{d\phi}{dt} = -\frac{\omega_0}{Q}\sin\beta,$$

the solution of this equation $\phi = -[\omega_0\sin\beta/Q]t + \phi_0$.

In this case, the time dependence of $E$ has the form $E = E_0\cos(\omega_0[1 + \sin\beta/Q]t - \phi_0)$; i.e., one can readjust the oscillation frequency of the VC device by varying the phase delay $\beta$. The maximal deviation is equal to $\Delta\omega = \omega_{max} - \omega_0 = 1/Q$. We see that the greater the $Q$-factor of the system, the smaller the frequency shift. Estimating the $Q$-factor of our cylindrical resonator (taking into account the inserted admittance due to approximation of the vector operators by finite-difference analogues [8]) we have obtained the value $Q \approx 10$, which agrees with the calculated maximal frequency shift.

To sum up, introducing the external delayed feedback into the VC system allows one to perform efficient control of the output characteristics of such devices.

## ACKNOWLEDGMENTS

This work was supported by the Russian Foundation for Basic Research, project no. 96-02-16753.